\documentclass[reprint, aps, pra, superscriptaddress, showpacs]{revtex4-1}

\usepackage{array}
\usepackage{graphicx}
\usepackage{amsmath}
\usepackage{amssymb}
\usepackage{hyperref}
\usepackage{color}
\usepackage{relsize}

\usepackage{txfonts}

\graphicspath{{figures/}}

\def\ba{\begin{eqnarray}}
\def\ea{\end{eqnarray}}
\def\beq{\begin{equation}}
\def\eeq{\end{equation}}

\newcommand{\OurNa}{~^{23}\mathrm{Na}}

\newcommand{\um}{~\mu\mathrm{m}}
\newcommand{\nm}{~\mathrm{nm}}

\newcommand{\Omegac}{\Omega_\textrm{c}}

\newcommand{\ket}[1]{|#1\rangle}
\newcommand{\bra}[1]{\langle#1|}

\newcommand{\NA}{\mathrm{NA}}

\newcommand{\xc}{w}
\newcommand{\wbarr}{\sigma}

\newcommand{\ketdark}{\ket{D(x)} }

\hypersetup{
    colorlinks=true,       
    linkcolor=blue,          
    citecolor=blue,        
    filecolor=blue,      
    urlcolor=blue           
}

\begin{document}

\title{Subwavelength-width optical  tunnel junctions for ultracold atoms}
\author{F.~Jendrzejewski}
\affiliation{Kirchhoff Institut f\"ur Physik, Universit\"at Heidelberg, Im Neuenheimer Feld 227, 69120 Heidelberg, Germany}
\author{S.~Eckel} 
\affiliation{Joint Quantum Institute, NIST/University of Maryland, College Park, MD 20742, USA}
\author{T.~G.~Tiecke}
\affiliation{Department of Physics, Harvard University, Cambridge, MA 02138, USA}
\affiliation{Facebook Inc., Connectivity Lab, 1 Hacker Way, Menlo Park, CA 94025, USA}
\author{G.~Juzeli\=unas}
\affiliation{Institute of Theoretical Physics and Astronomy, Vilnius University, Saul\.{e}tekio ave.\ 3, LT-10222 Vilnius, Lithuania}
\author{G.~K.~Campbell}
\affiliation{Joint Quantum Institute, NIST/University of Maryland, College Park, MD 20742, USA}
\author{Liang Jiang}
\affiliation{Departments of Applied Physics and Physics, Yale University, New Haven, CT 06520, USA}
\author{A.~V.~Gorshkov}
\affiliation{Joint Quantum Institute, NIST/University of Maryland, College Park, MD 20742, USA}
\affiliation{Joint Center for Quantum Information and Computer Science, NIST/University of Maryland, College Park, MD 20742, USA}

\begin{abstract}
We propose a new method for creating far-field optical barrier potentials for ultracold atoms with widths that are narrower than the diffraction limit and can approach tens of nanometers. The reduced widths stem from the nonlinear atomic response to control fields that create  spatially varying dark resonances. The subwavelenth barrier is the result of the geometric scalar potential experienced by an atom prepared in such a spatially varying dark state. The performance of this technique, as well as its applications to the study of many-body physics and to the implementation of quantum information protocols with ultracold atoms, are discussed, with a focus on the implementation of tunnel junctions.
\end{abstract}

\maketitle

\section{Subwavelength-width barrier}

Optical dipole potentials are an essential tool for  the manipulation of degenerate quantum gases. For example, they provide flexible trapping geometries \cite{Grimm2000}, highly tunable optical lattices \cite{Bloch2005}, and well-controlled disordered potentials \cite{Modugno2010}.  However, unless operated near surfaces \cite{Chang2009,gullans12,romero-isart13,Thompson2013,Gonzalez-Tudela2015,Mitsch2014}, spatial features of such potentials are typically limited by diffraction to be wider than half of an optical wavelength  $\lambda$.
The use of a three-level system to generate an optical potential in place of the typical two-level system  offers more flexibility, as dark resonances \cite{scully1997quantum} allow one to overcome the diffraction limit \cite{gorshkov08,bajcsy03,dutton01,ginsberg05,sahrai05,agarwal06b,cho07,yavuz07,juzeliunas07,Kiffner2008,miles13}. 
In this paper, building on previous 
studies of subwavelength-scale forces \cite{prentiss91}, atom localization 
\cite{Kiffner2008,kapale13,johnson98,thomas89,stokes91,schrader04,gardner93,zhang06,lee07,bajcsy03,dutton01,ginsberg05,sahrai05,agarwal06b,kapale10,le-kien97,qamar00,paspalakis01,hell07,yavuz07,cho07,juzeliunas07,gorshkov08,li08e,mompart09,sun11,proite11,qi12b,viscor12,yavuz12,rubio13,miles13}, and non-dark-state-based techniques for building subwavelength potentials in the far field \cite{Yavuz2009, yi08, shotter08, lundblad08, gupta96,zhang05b,berman98,dubetsky02, weitz04,Nascimbene2015,Brezger1999,Salger2007},  
we use the geometric scalar (Born-Huang) potential \cite{mead92,dum96,goldman14} experienced by spatially dependent dark states  
to create optical potential barriers with subwavelength widths.   
  Our proposal has the advantage of not using lattice modulation, which could lead to heating, and of taking advantage of a feature -- the geometric scalar potential -- that naturally accompanies any subwavelength potential formed by spatially dependent dressed states. Furthermore, the use of dark states contributes to reduced spontaneous emission in our scheme.

\begin{figure}[b]
\begin{center}
\includegraphics[width = 0.99 \columnwidth]{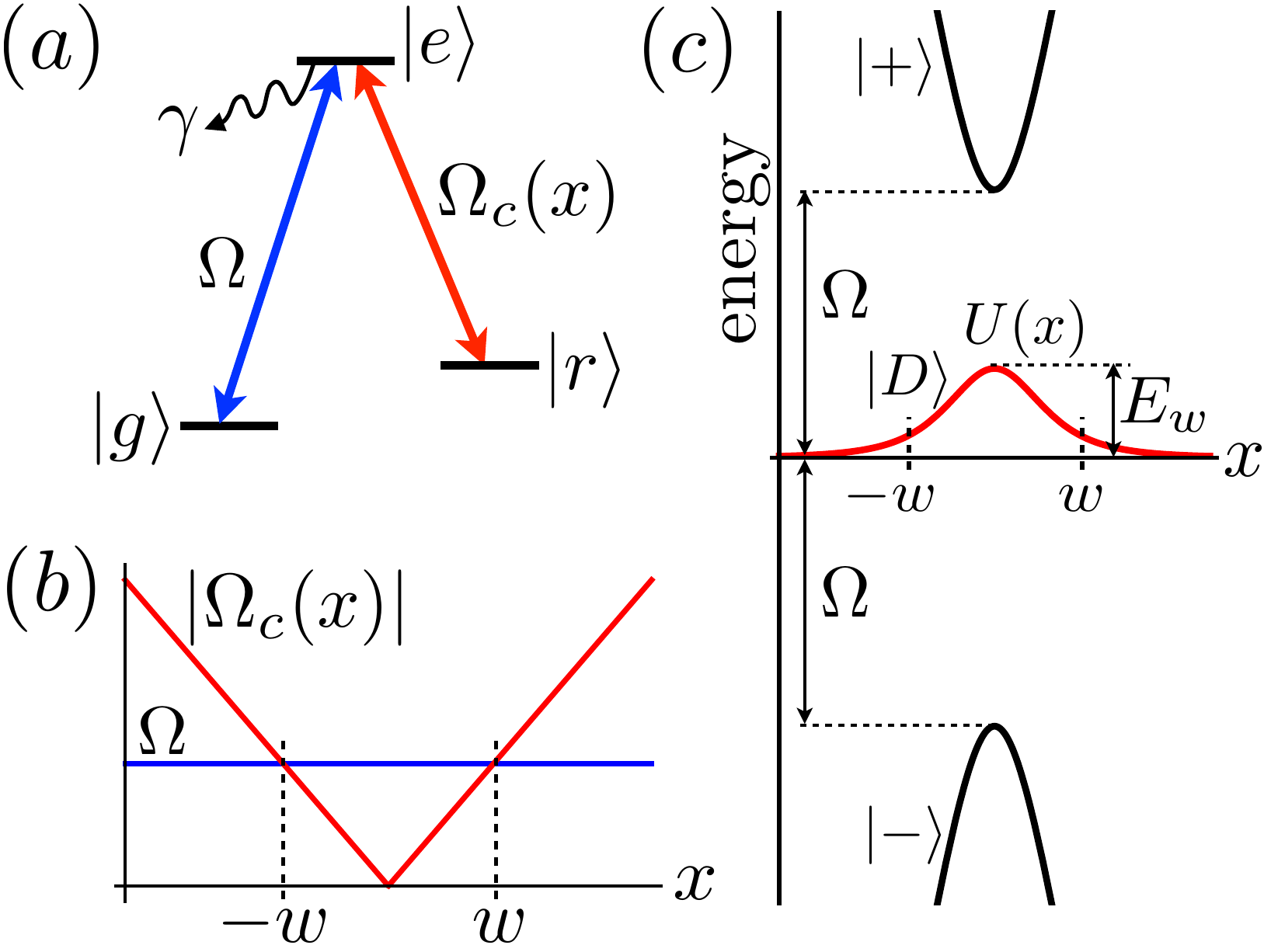}
\caption{The scheme for creating a subwavelength barrier. (a) The level diagram featuring a resonant probe field with a spatially uniform Rabi frequency $\Omega$ and a resonant control field with spatially-dependent Rabi frequency $\Omegac(x)$. $\gamma$ is the linewidth of the excited state. (b) The spatial dependence of the probe and control Rabi frequencies. (c) The geometric scalar potential $U(x)$ of the dark state $\ket{D}$ features a potential barrier of height $E_w = 1/(2 m w^2)$ and subwavelength width $\xc$ determined by $\Omega = \Omegac(\xc)$. The energies of the other two rotating-frame eigenstates $\ket{+}$ and $\ket{-}$ are shown schematically. 
 \label{fig:scheme}} 
\end{center}
\end{figure}

The main idea of the proposed scheme can be understood using the three-state model shown in Fig.~\ref{fig:scheme}(a). A probe field with a spatially uniform Rabi frequency $\Omega$ resonantly couples states $\ket{g}$ and $\ket{e}$. A control field resonantly couples states $\ket{r}$ and $\ket{e}$ and has Rabi frequency $\Omegac(x)$ that depends on position $x$, as shown in Fig.\ \ref{fig:scheme}(b). While far from the origin this control field is much stronger than the probe field ($\Omegac \gg \Omega$), the control field vanishes at $x = 0$. For zero 
detuning, $\delta=0$, destructive interference of excitation pathways from $\ket{g}$ and $\ket{r}$ up to $\ket{e}$ ensures that the so-called ``dark state'' $\ketdark = \left[\Omega \ket{r}-\Omegac(x) \ket{g}\right]/\sqrt{\Omega^2+\Omegac^2(x)}$ is decoupled from both optical fields  and has zero energy in the frame rotating with the applied optical fields \cite{scully1997quantum}.
To transform to a new basis, where $\ketdark$ is one of the basis states, we apply a spatially dependent unitary transformation, which acts non-trivially on the kinetic-energy (since the latter involves spatial derivatives) and, as we will show in detail below [see Eq.\ (\ref{eq:HV})], endows the dark state with a geometric scalar potential ($\hbar = 1$) \cite{mead92,dum96,goldman14}
\begin{eqnarray}\label{eq:geom}
U(x) = \frac{1}{2 m} \left(\frac{\Omega \, \partial_x \Omega_c(x)}{\Omega^2+\Omegac^2(x)}\right)^2 =  \frac{E_w}{(1 + (x/w)^2)^2},
\ea 
where, in the last equality, we assumed for concreteness a linear position dependence of the Rabi frequency, $\Omega_c(x) = \Omega s x/\sigma$, which is valid around the node. Here $\sigma \sim \lambda$ is a diffraction-limited length scale, the dimensionless parameter $s$ controls the amplitude of the control field relative to the probe field, $w = \sigma/s$ is the width of the resulting potential, $E_w = 1/(2 m w^2)$ is the characteristic kinetic energy associated with this width, and $m$ is the mass of the atom. As shown in Fig.\ \ref{fig:scheme}(c), this creates a potential barrier of height $E_w$ and width $\xc$ determined via $\Omega = \Omegac(\xc)$.
Stated simply, the dark state is $\ket{g}$ for $\Omegac(x) \gg \Omega$, while $\ket{r}$ is the dark state for $\Omegac(x) \ll \Omega$.  As an atom moves along $x$, the change of the internal state from $\ket{g}$ to $\ket{r}$ and back happens over the short lengthscale $\xc$. The geometric scalar potential $U(x)$ provides a potential barrier that effectively describes 
the kinetic energy of atomic micromotion \cite{aharonov92,cheneau08} emerging due to the rapid change of the internal state in this region. 
Crucially, the width of the barrier is  limited not by an optical wavelength but by the ratio of $\Omega$ and the slope of $\Omegac(x)$, which in turn depends on the control beam intensity. The height $E_w$ of the potential is determined by the width $w$ and, due to the inverse quadratic scaling, makes the barrier impenetrable for $w \rightarrow 0$ while keeping incoming energy $E$ fixed.

\begin{figure}[b]
\begin{center}
\includegraphics[width = 1 \columnwidth]{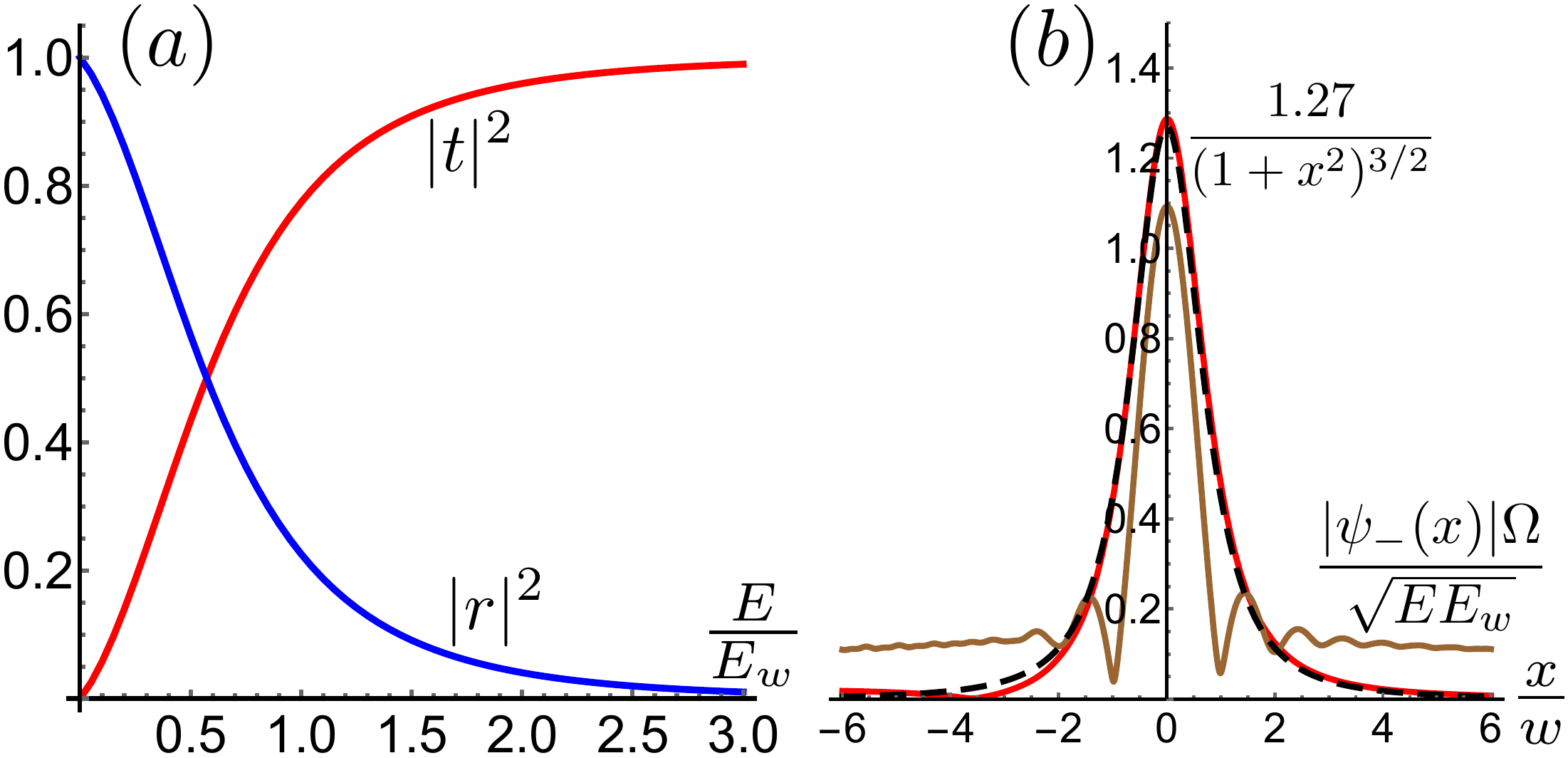} 
\caption{\label{fig:2} (a) Transmission $|t|^2$ and reflection $|r|^2$ probabilities for the geometric scalar potential $U(x)$ [Eq.\ (\ref{eq:geom})] as a function of the incoming energy $E$. (b) $|\theta'' \psi_D + 2 \theta' \psi'_D| \sqrt{E_w/E}$ (red), its approximate form $1.27/(1+x^2)^{3/2}$ from  
 Eq.\ (\ref{eq:g}) (black dashed), and $|\psi_-(x)| \Omega/\sqrt{E E_w} $ (brown) for $\Omega = 20 E_w$ and $E = E_w/50$. Excited state population can be determined via $P_e \sim |\psi_-(0)|^2 \approx  
 \frac{E E_w}{\Omega^2}$, which agrees with Eq.\ (\ref{eq:pe}), while the nonadiabaticity error is $P_\textrm{NA} = \frac{k_2}{k} \left(|\psi_-(-L)|^2 + |\psi_-(L)|^2\right) \approx  
 9 \times 10^{-5}$, which agrees with the approximate asymptotic formula [Eq.\ (\ref{eq:pna})] $p_\textrm{NA} \approx 1.37 \sqrt{E/E_w}  e^{- 1.75 \sqrt{\Omega/E_w}} \approx 8 \times 10^{-5}$.}   
\end{center}
\end{figure}

If one rescales $x$ by $w$, one sees that the eigenvalue problem in the presence of the potential in Eq.\ (\ref{eq:geom}) depends only on the dimensionless parameter $E/E_w$.  
Transmission $|t|^2$ and reflection $|r|^2$ probabilities are shown as a function of $E/E_w$ in Fig.\ \ref{fig:2}(a). For our applications, in which particles either tunnel through the barrier or use the barrier as a wall, we will be interested in $E < E_w$. It is therefore useful to evaluate the transmission probability through the barrier in the limit $E/E_w \rightarrow 0$, which we numerically find to scale linearly with $E/E_w$ and be approximately given by $|t|^2 \approx 0.4 E/E_w$. Below, we will use our numerical solution for the ideal problem of scattering from $U(x)$ to analyze imperfections and specific applications of the barrier.  

The ability to generate subwavelength barriers opens numerous avenues of research with ultracold atoms. First, since these barriers become impenetrable at fixed $E$ as 
$w \rightarrow 0$,
they can be used to create sharp walls with a subwavelength rise distance. 
One application of such sharp walls is the creation of a two-dimensional or three-dimensional optical-box trap \cite{Gaunt2013a}, where sharp walls are important, for example, for defining and detecting topological edge states \cite{goldman10b,Goldman2013}. One could also make a one-dimensional optical-box trap, where the other two dimensions are frozen out. Treating occupied orbitals of such a square-well potential as sites, one can realize highly symmetric spin models 
\cite{beverland16} and do efficient spectrum estimation of density operators \cite{beverland16b}. As discussed in the supplemental material to Ref.\ \cite{beverland16}, sharp walls are important for these applications. 
 Second, introducing subwavelength features into a disordered potential may 
modify the percolation properties of such a disorder 
compared to the widely used speckle disorder \cite{Robert-de-Saint-Vincent2010} and may open new avenues in the study of Anderson localization with matter waves. 

Third, subwavelength barriers are great candidates for a robust implementation of very narrow tunnel junctions. Let $\xc$ be the junction width and $\xi$ be either the healing length of the condensate or the deBroglie wavelength of a thermal cloud. 
As long as $\xi < \xc$, the dynamics can be  
described using the local density approximation. On the other hand, when $\xi \gg \xc$, one reaches the tunnel-junction limit. In a typical BEC experiment, the condensate healing length is $\xi \approx 0.5$ $\mu$m \cite{Eckel2014}. While it is, in principle, possible to increase $\xi$, it is a demanding experimental task as it requires working with extremely dilute clouds, having good optical access (high numerical aperture)  \cite{Albiez2005, Ryu2013a}. Additionally, the scattering length could be tuned through a suitable Feshbach resonance in certain atomic spieces like \textsuperscript{39}K or \textsuperscript{7}Li. However, this tool would be challenging to use for the widely employed \textsuperscript{23}Na and \textsuperscript{87}Rb. Further, Feshbach resonances are only available at very specific magnetic bias fields such that they are larglely incompatible with the growing field of spinor condensates \cite{Stamper-Kurn2013}. 
Therefore, subwavelength techniques are highly desirable for achieving the tunnel-junction limit.  
Tunnel junctions may also enable  the realization of Josephson effects \cite{Adhikari2009a} or Andreev bound states \cite{Zapata2009,VanSchaeybroeck2007}.

To be specific, we will focus in this manuscript on the implementation of  tunnel junctions for a ring BEC,  which will open avenues for a wide variety of  studies. 
A lattice of tunnel junctions on a ring \cite{Dziarmaga2011} (created, for example, by superposing control beams with opposite vorticity) will enable  the creation of lattice  models with periodic boundary conditions and with the resulting ability to support  piercing magnetic fluxes and the associated topological phenomena. 
(Lattice spacings smaller than $\sim$$10$ $\mu$m on a ring are challenging to achieve \cite{Amico2014}; without using subwavelength techniques, tunneling rates in such a lattice will be negligible.)

\section{Imperfections}

For the remainder of the paper, we will study deviations from Eq.\ (\ref{eq:geom}). While we will focus for simplicity on a one-dimensional setup, extensions to higher-dimensional barriers are straightforward. In the rotating frame, under the rotating-wave approximation, and ignoring the decay rate of state $\ket{e}$,  the Hamiltonian of a single atom in the \{$\ket{r}$, $\ket{g}$, $\ket{e}$\} basis  is 
\begin{eqnarray}
H = \frac{p^2}{2 m} +  \left (\begin{array}{ccc}
0 & 0 &  \Omegac(x) \\
0 & 0 &  \Omega \\
 \Omegac(x) &  \Omega & 0
\end{array}\right).
\end{eqnarray}
Destructive interference produces the dark eigenstate  $\ket{D(x)} = (\Omegac(x) \ket{g} - \Omega \ket{r})/\overline \Omega(x)$ that, in the absence of $p^2/(2 m)$, is decoupled from the light fields and has an eigenvalue of zero, where $\overline \Omega(x) = \sqrt{\Omega^2 + \Omegac^2(x)}$. The orthogonal bright state in the $\{\ket{r},\ket{g}\}$ subspace, $\ket{B(x)} = (\Omegac(x) \ket{r} + \Omega \ket{g})/\overline \Omega(x)$, then couples resonantly to state $\ket{e}$ with Rabi frequency $\overline \Omega(x)$, giving rise to eigenstates $\ket{\!\pm\!(x)} = (\ket{B(x)} \pm \ket{e})/\sqrt{2}$ with energies $\pm \overline \Omega(x)$. To change the Hamiltonian into  the $\{\ket{D}, \ket{+},\ket{-}\}$ basis, it is convenient to define the unitary operator 
\ba
R^\dagger = \ket{\tilde D}\bra{D(x)} + \ket{\tilde +}\bra{+(x)} + \ket{\tilde -}\bra{-(x)},
\ea
where $\ket{\tilde D}$ and $\ket{\tilde \pm}$ are position-independent vectors, which can be thought of as $\ket{D(x_0)}$ and $\ket{\!\pm\!(x_0)}$ evaluated at some fixed position $x_0$. In the $\{\ket{\tilde D}, \ket{\tilde +},\ket{\tilde -}\}$ basis, the Hamiltonian is 
 \begin{eqnarray}\label{eq:diag}
H' = R^\dagger H R &=&  \frac{(p-A)^2}{2 m}  + \sum_{a = \pm} a \overline \Omega(x) \ket{\tilde a}\bra{\tilde a},
\end{eqnarray}
where the effective vector potential  $A= i R^\dagger \partial_x R$ is \cite{goldman14}
\ba
A = i \frac{\Omega \, \partial_x \Omega_c(x)}{\overline \Omega^2(x)}  \left(\ket{\tilde B} \bra{\tilde D} - \ket{\tilde D} \bra{\tilde B}\right)
\ea
 and $\ket{\tilde B} = (\ket{\tilde +} + \ket{\tilde -})/\sqrt{2}$. Therefore,
 \ba
\!\!\!\! H' = \frac{p^2}{2 m} + U(x) \ket{\tilde D}\bra{\tilde D} + \sum_{a = \pm} \left(a \overline \Omega(x) + \frac{U(x)}{2}\right) \ket{\tilde a}\bra{\tilde a} + V, \label{eq:HV}
 \ea
 where $U(x) = \bra{\tilde D} A^2 \ket{\tilde D}/(2 m)$ is the geometric scalar potential given by Eq.\ (\ref{eq:geom}). In Eq.\ (\ref{eq:HV}), we have explicitly presented the terms diagonal in the $\{\ket{\tilde D}, \ket{\tilde +},\ket{\tilde -}\}$ basis. An additional off-diagonal contribution
 \ba
 V = - \frac{p A + A p}{2m} + \frac{U(x)}{2} (\ket{\tilde +}\bra{\tilde -} + \ket{\tilde -}\bra{\tilde +})
\ea
describes the coupling between these basis states. Since $V \sim E_w$, we should be able to treat $V$ perturbatively in the limit $\Omega \gg E_w$, in which $\ket{\tilde D}$ is separated from $\ket{\tilde \pm}$ by $\approx \Omega$ at the distance of closest approach. We will assume this limit, as well as $E \ll E_w$ for the remainder of the manuscript.

We now analyze the two main imperfections associated with the subwavelength barrier. The first imperfection is the nonzero probability $P_\textrm{NA}$ of losing the atom into the open $\ket{\tilde -}$ channel due to the non-adiabatic coupling between $\ket{\tilde D}$ and $\ket{\tilde -}$. Since the non-adiabatic coupling is $\sim E_w$ and the minimal gap to $\ket{\tilde -}$ is $\Omega$, we expect $P_\textrm{NA}$ to be small provided the adiabaticity condition $\Omega \gg E_w$ is satisfied. We will derive this adiabaticity condition below.

The second imperfection is the nonzero probability $P_\textrm{sc}$ of spontaneously scattering a photon from the excited state $\ket{e}$, which can be populated due to the non-adiabatic coupling between $\ket{\tilde D}$ and $\ket{\tilde \pm}$, which in turn have $\ket{e}$ components. Such a scattering event can lead to heating and loss of atoms as a single scattered photon can transfer the atom out of the dark state $\ket{\tilde D}$.  We will work in the limit $\Omega \gg \gamma$, where spontaneous emission can be analyzed perturbatively: we will calculate the typical probability $P_e$ of the excited state at $x=0$ by solving the scattering problem in the absence of spontaneous emission, and then use that to evaluate the probability of scattering a photon in a single pass  
 $P_\textrm{sc} \sim \gamma P_e (w/\sqrt{E/m})$, where the term in parentheses is the crossing time.

It is easy to estimate on physical grounds the probability $P_e$ of the excited state at $x=0$. In the original units, the dark state probability at $x=0$ is $\sim |t|^2 \sim E/E_w$. The admixture of $\ket{\tilde -}$ (and hence of $\ket{e}$) is simply $\sim (E_w/\Omega)^2$, where the non-adiabatic coupling $E_w$ plays the role of an effective Rabi frequency and $\Omega$ plays the role of detuning. Multiplying the two together we obtain
\ba
P_e \sim \frac{E E_w}{\Omega^2}, \label{eq:pe}
\ea
which agrees well with the more precise calculation below.  The probability of scattering a photon in a single pass is then $P_\textrm{sc} \sim \gamma \sqrt{E E_w}/\Omega^2$.

We now evaluate the imperfections more precisely for $\Omega_c(x) = \Omega s x/\sigma = \Omega x/w$. Measuring distances in units of $w$ and energies in units of $E_w$, the equations become independent of $s$. In particular, we are looking for an eigenstate of energy $E$ of the form $\ket{\psi} = \psi_g \ket{g} + \psi_e \ket{e} + \psi_r \ket{r}$ that has an incoming dark-state plane wave from the left. This energy $E$ and the probe Rabi frequency $\Omega$ are the only two remaining parameters in the problem. The eigenvalue equations are
\ba
E \psi_g &=& - \psi_g'' + \Omega \psi_e, \label{eq:psig}\\
E \psi_e &=& - \psi_e'' + \Omega \psi_g + \Omega_c \psi_r,\\
E \psi_r &=& - \psi_r'' + \Omega_c \psi_e \label{eq:psir},
\ea
where $\Omega_c = \Omega x$, and $E$ and $\Omega$ have been rescaled by $E_w$. We define the incoming $k$ vector (in units of $w$) as $k = \sqrt{E}$ and the $k$ vector in the $\ket{\tilde -}$ channel at $x = L \gg 1$ far away from the barrier as $k_2 = \sqrt{\Omega L}$.   
Defining the $\ket{\tilde D},\ket{\tilde B},\ket{\tilde \pm}$ amplitudes as
\ba
\psi_D &=& \frac{x \psi_g - \psi_r}{x^2 + 1},\\
\psi_B &=& \frac{x \psi_r + \psi_g}{x^2 + 1},\\
\psi_\pm &=&  \frac{\psi_B \pm \psi_e}{\sqrt{2}},
\ea
the boundary conditions at $-L$ and $+L$  are 
\ba
\psi'_D(L) &=& i k \psi_D(L), \label{eq:b1}\\
\psi'_D(-L)  &=&-  i k \psi_D(-L) + 2 i k e^{- i k L},\\
\psi_+(\pm L) &=& 0,\\
\psi'_-(\pm L) &=& \pm i k_2 \psi_-(\pm L). \label{eq:b2}
\ea
These correspond, respectively, to: incoming $\ket{\tilde D}$ from the right vanishes; incoming $\ket{\tilde D}$ from the left has unit amplitude; $\ket{\tilde +}$ is unoccupied at $\pm L$; 
incoming $\ket{\tilde -}$ from the right/left vanishes. 

The transmission and reflection amplitudes of the dark state are then 
\ba
t &=& \psi_D(L) e^{- i k L},\\
r &=& (\psi_D(-L) - e^{- i k L}) e^{-i k L}.
\ea
We then define the non-adiabaticity error as 
\ba
p_\textrm{NA} = 1 - |t|^2 - |r|^2 = \frac{k_2}{k} \left(|\psi_-(-L)|^2 + |\psi_-(L)|^2\right),
\ea
which is the probability that, instead of staying in the $\ket{\tilde D}$ channel, the outgoing particle leaves in the open $\ket{\tilde -}$ channel. 

To simplify the problem, we start by changing the basis from $\{\ket{g},\ket{e},\ket{r}\}$ to $\{\ket{\tilde D}, \ket{\tilde +},\ket{\tilde -}\}$. The resulting equations of motion are
\ba
\!\!\!\!\! \!\!\! E \psi_D &=& - \psi_D'' + U \psi_D + \tfrac{1}{\sqrt{2}} (\theta'' (\psi_+ + \psi_-) + 2 \theta' (\psi_+' + \psi_-')),\\
\!\!\!\!\! \!\!\! E \psi_+ &=& - \psi_+'' + \overline \Omega \psi_+ + \tfrac{U}{2} (\psi_+ + \psi_-) - \tfrac{1}{\sqrt{2}} (\theta'' \psi_D + 2 \theta' \psi_D'),\\
\!\!\!\!\! \!\!\! E \psi_- &=& - \psi_-'' - \overline \Omega \psi_- + \tfrac{U}{2} (\psi_+ + \psi_-) - \tfrac{1}{\sqrt{2}} (\theta'' \psi_D + 2 \theta' \psi_D'),
\ea
where $U = 1/(1+x^2)^2$, $\theta = \arctan x$, and $\overline \Omega = \sqrt{x^2 + 1}$ in the rescaled units.

Since $\ket{\tilde +}$ is a closed channel, for the purpose of calculating the adiabaticity error  and the population $P_e$ of $\ket{e}$, it is sufficient to simply set $\psi_+ = 0$. We have checked numerically that, within our regime of interest ($\Omega \gg 1$ and $E \ll 1$), this is an excellent approximation. Furthermore, in this limit, for the purposes of calculating $|\psi_-(\pm L)|^2$ and $P_e$, it is an excellent approximation to drop all off-diagonal terms from the equation for $\psi_D$. Since $E \ll 1$ and $\Omega \gg 1$, we can also drop $E \psi_-$ and $U \psi_-/2$ from the $\psi_-$ equation. This leaves us with
\ba
E \psi_D &=& - \psi_D'' + U \psi_D,\\
0 &=& - \psi_-'' - \overline \Omega \psi_- - \frac{1}{\sqrt{2}} (\theta'' \psi_D + 2 \theta' \psi_D'). \label{eq:psim2}
\ea
The equation for $\psi_D$ is now just the ideal single-channel equation whose solution we have already obtained numerically above. Specifically, as we show in Fig.\ \ref{fig:2}(b), in the limit $E \ll 1$,  the following is an excellent approximation
\ba
\theta'' \psi_D + 2 \theta' \psi_D' \approx i \frac{1.27 \sqrt{E}}{(1 + x^2)^{3/2}}. \label{eq:g}
\ea 
The resulting simple second-order differential equation for $\psi_-$ [Eq.\ (\ref{eq:psim2})] can be solved using variation of parameters, where we construct the actual solution from two general solutions of the homogeneous equation. Furthermore, for the solution to the homogeneous equation, we use the WKB approximation. Referring the reader to the Appendix for straightforward details of this solution, here we simply present the answers and show in Fig.\ \ref{fig:2}(b) the numerical calculation of $|\psi_-(x)|$ using Eqs.\ (\ref{eq:psig}-\ref{eq:psir}) and Eqs.\ (\ref{eq:b1}-\ref{eq:b2}).  In particular, we find in the Appendix and confirm numerically in Fig.\ \ref{fig:2}(b) that
\ba
P_e \sim |\psi_-(0)|^2 \sim E/\Omega^2, \label{eq:pe2}
\ea
which recovers Eq.\ (\ref{eq:pe}), as desired.

Furthermore, we find in the Appendix and confirm numerically in Fig.\ \ref{fig:2}(b) that $|\psi_-(L)| = |\psi_-(-L)|$ and that
\ba
p_\textrm{NA} = 2 \frac{k_2}{k}  |\psi_-(L)|^2 \approx 1.37 \sqrt{E}  e^{- 1.75 \sqrt{\Omega}}. \label{eq:pna}
\ea
As predicted, $p_\textrm{NA} \rightarrow 0$ as $\Omega \rightarrow \infty$. Furthermore, $p_\textrm{NA} \rightarrow 0$ as $E \rightarrow 0$ since, in this limit, the incoming particle moves very slowly allowing for near-perfect adiabaticity) and does not penetrate significantly into the barrier. 

\section{Experimental parameters}
 
We now estimate the relevant experimental parameters for an implementation of tunnel junctions for a BEC of $\OurNa$, using for concreteness the configuration of Ref.~\cite{Eckel2014}. We assume the control field has an intensity profile  
proportional to $\Omegac^2(x) = \Omega^2 s^2(1-e^{-x^2/\wbarr^2})$, 
so that $\Omegac^2(x) \simeq \Omega^2 \frac{s^2 x^2}{\wbarr^2}$ for $x \ll \sigma$.  
Such an intensity profile can  be approximately implemented by using existing techniques  
\cite{Eckel2014,Gaunt2013a}. 
For a given numerical aperture $\NA$, the spatial scale $\sigma$ and hence the smallest feature of such an intensity profile is constrained by the diffraction limit to be greater than or equal to $\sigma = \frac{\lambda}{\sqrt{2}\pi\NA}$ \cite{hecht2002optics}. We will take $\sigma = 3 \um$.  For an incoming energy equal to the chemical potential, i.e.\ $E = 2 \pi \times 1\textrm{ kHz} \approx 1/(2 m \xi^2)$, without using subwavelength techniques, a barrier potential $V_0 e^{-x^2/\wbarr^2}$ would exhibit appreciable tunneling tunneling ($|t|^2 > 0.01$) only for a narrow window $E < V_0 < 1.25 E$, making tunneling effects negligible in the experiment of Ref.\ \cite{Eckel2014}. To increase the effects of tunneling, we choose the desired barrier width $\xc = 100\nm$, which gives $s = \sigma/\xc =30$ and barrier height $E_w = 2 \pi \times 22$ kHz, corresponding to $|t|^2 \approx 0.02$ for an incoming energy $E = 2 \pi \times 1\textrm{ kHz}$. We see that, for these parameters, tunneling is observable even for an incoming energy that is a factor of 20 smaller than the barrier height, making tunneling crucial in the transport of atoms through such links.  
Taking the maximum available control Rabi frequency to be $\Omega s  = 2 \pi\times 200$ MHz, we have $\Omega = 2 \pi \times 6.7$ MHz, so the adiabaticity condition $\Omega \gg E_w$ is satisfied. Since we would like to be in the tunneling regime $E < E_w$, using $\gamma = 2 \pi\times 10$ MHz, we find that the photon scattering probability $P_\textrm{sc} = \gamma \sqrt{E E_w}/\Omega^2 < \gamma E_w/\Omega^2 = 0.005$ is much smaller than one, as desired \footnote{Strictly speaking, our derivation of $P_\textrm{sc}$ assumes $\Omega \gg \gamma$, however we expect this estimate to still be qualitatively correct at these values of $\Omega \sim \gamma$.}. The implementation of the barrier will require the residual control amplitude at the node to be at most $s$ times smaller than the maximum control field amplitude, which is easily achievable.

The adiabaticity condition and the maximum achievable control Rabi frequency limit the smallest achievable $w$ to $> [\sigma/(2 m \Omega s)]^{1/3} = 15$ nm, corresponding to $s = 200$ and $\Omega = E_w = 2 \pi \times 1$ MHz. A barrier of such height would be nearly impenetrable [$|t|^2 < 4 \times 10^{-4}$]  for typical incoming energies [$< (2 \pi)1$ kHz] and would thus function as a sharp wall. For an incoming kinetic energy of $(2 \pi) 1$ kHz, the photon scattering probability $P_\textrm{sc} \approx 0.3$ is still modest, as desired. However, this estimate of $P_\textrm{sc}$ likely has to be modified to take into account the regime $\Omega \ll \gamma$.  

\section{Outlook}

While we have focused on the case where both applied fields are on resonance with the corresponding transition, the introduction of single-photon and two-photon detuning may provide additional flexibility for improving the performance of the barrier. 
The technique we presented allows for the generation of an array of barriers that have subwavelength widths but whose separation is still diffraction-limited. By stroboscopically shifting the resulting potential to different positions \cite{Nascimbene2015}, one might be able to create a time-averaged potential where barriers of subwavelength width are separated by subwavelength distances. Alternatively, subwavelength distances between barriers can be achieved by using additional levels and additional beams, with $n$ beam pairs (i.e.\ $\Omega$ and $\Omega_c$) required to divide the lattice constant by $n$. Lattices with reduced lattice constants 
will allow for the realization of Hubbard-type models with increased tunneling and interaction energies 
\cite{gullans12,romero-isart13,Gonzalez-Tudela2015,yi08,lundblad08}.
Such increased energy scales will reduce temperature and coherence requirements for studying these models.

\textit{Note.}---While completing the manuscript, we learned of a related 
proposal for designing  
subwavelength barriers with a focus on creating double-layer potentials \cite{lacki16}.

\begin{acknowledgements}

We thank T.\ Calarco, M.\ Lukin, J.\ Thompson, T.\ Porto, S.\ Rolston, W.\ Phillips, I.\ Spielman, P.\ Julienne, E.\ Tiesinga, and R.\ Qi for discussions. This work was partially supported by the NSF PFC at JQI, AFOSR, NSF QIS, ARO, ARO MURI, and ARL CDQI. G.J. was supported by the Lithuanian Research Council (Grant No.~MIP-086/2015).
\end{acknowledgements}

\appendix*

\section{Solution of Eq.\ (\ref{eq:psim2})}

Here we solve the second-order differential equation for $\psi_-$ [Eq.\ (\ref{eq:psim2})], which allows us to derive the expressions for the excited state population $P_e$ inside the barrier [Eq.\ (\ref{eq:pe2})] and the probability $P_\textrm{NA}$ of the non-adiabatic loss of the atom into the $\ket{\tilde -}$ channel [Eq.\ (\ref{eq:pna})].

We solve the second-order differential equation for $\psi_-$ [Eq.\ (\ref{eq:psim2})] using variation of parameters, where we construct the actual solution from two general solutions of the homogeneous equation. Furthermore, for the solution to the homogeneous equation, we use the WKB approximation. The answer is then computed as follows. These are the two homogeneous solutions in the WKB approximation: 
\ba
f_1(x) &=& \left(\frac{L^2 + 1}{x^2 + 1}\right)^{1/8} e^{- i \sqrt{\Omega} \int_{-L}^x (y^2 + 1)^{1/4} d y},\\
f_2(x) &=& \left(\frac{L^2 + 1}{x^2 + 1}\right)^{1/8} e^{- i \sqrt{\Omega} \int_x^L (y^2 + 1)^{1/4} d y}.
\ea
We chose the normalization so that $f_1(-L) = f_2(L) = 1$. Calculating the Wronskian, we find that, under the WKB approximation, it is independent of $x$:
\ba
W = f_1 f_2' - f_1' f_2 = - i 2 \sqrt{\Omega} (L^2 + 1)^{1/4}.
\ea
The solution for $\psi_-$ is then
\ba
 \!\!\!\!\!\!\!\!\!\!\!  \psi_-(x) = -f_1(x) \int_x^L \! \frac{f_2(y) g(y)}{W} d y - f_2(x) \int_{-L}^x \! \frac{f_1(y) g(y)}{W} d y,
\ea
where we used Eq.\ (\ref{eq:g})  to define 
\ba
g(x) = \frac{1}{\sqrt{2}} (\theta'' \psi_D + 2 \theta' \psi_D') \approx  i \frac{0.90 \sqrt{E}}{(1 + x^2)^{3/2}}.
\ea

It is easy to check that 
\ba
\int_0^L f_2(y) g(y) d y \sim \int_{-L}^0 f_1(y) g(y) \sim L^{1/4} \sqrt{E/\Omega},
\ea
which immediately yields $P_e$ in Eq.\ (\ref{eq:pe2}).

We also see that $|\psi_-(L)| = |\psi_-(-L)|$, so that
\ba
\!\!\!\! \!\!\!\! \!\!\!\! p_\textrm{NA} &=& 2 \frac{k_2}{k}  |\psi_-(L)|^2 = 2 \frac{k_2}{k} \left|\int_{-L}^L \frac{f_1(x) g(x)}{W} d x\right|^2 \\
\!\!\!\! \!\!\!\! \!\!\!\!  &=& \frac{0.41 \sqrt{E}}{\sqrt{\Omega}}  \left|\int_{-L}^L (1+x^2)^{-13/8} e^{- i \sqrt{\Omega} \int_0^x (y^2+1)^{1/4} dy} d x\right|^2.
\ea
To take the $\Omega$-dependent integral inside the norm in the limit $\Omega \gg 1$,  
we expand the exponent around $x = - i$,
\ba
\!\!\!\! \!\!\!\! i \int_0^x  (y^2+1)^{1/4} dy \approx \frac{\pi^{3/2}}{3 \sqrt{2} \Gamma^2(3/4)} - \frac{2^{1/4} 4 }{5} (1 - i x)^{5/4}, 
\ea 
where $\Gamma$ is the Gamma function, so that
\ba
\!\!\!\! \!\!\!\! \!\!\!\! p_\textrm{NA} \approx \frac{0.41 \sqrt{E}}{\sqrt{\Omega}}  e^{- \sqrt{\Omega} \frac{\sqrt{2} \pi^{3/2} }{3  \Gamma^2(3/4)}} \left[\int_{-L}^L  \frac{e^{\sqrt{\Omega} \frac{2^{1/4} 4}{5} (1 - i x)^{5/4}}}{(1+x^2)^{13/8}} d x\right]^2.
\ea
We have therefore extracted the most sensitive exponential dependence on $\sqrt{\Omega}$. Numerics show that, possibly up to small corrections, the remaining integral at large $\Omega$ is given by
 \ba
 \int_{-L}^L  \frac{e^{\sqrt{\Omega} \frac{2^{1/4} 4}{5} (1 - i x)^{5/4}}}{(1+x^2)^{13/8}} d x \approx 1.84 \Omega^{1/4}.
 \ea
 The final result is therefore
 \ba
p_\textrm{NA} \approx 1.37 \sqrt{E}  e^{- \sqrt{\Omega} \frac{\sqrt{2} \pi^{3/2} }{3  \Gamma^2(3/4)}},
\ea
which yields Eq.\ (\ref{eq:pna}).

\end{document}